\documentclass[preprint,11pt]{elsarticle}
\usepackage[top=1in,bottom=1in,right=1in,left=1in]{geometry}
\usepackage[usenames, dvipsnames]{color}
\definecolor{g}{RGB}{0, 150, 0}
\usepackage{url,epsfig,amsmath,paralist}
\usepackage[ruled,vlined]{algorithm2e}
\SetKwInOut{KwInit}{Initialization}

\usepackage[top=1in,bottom=1in,right=1in,left=1in]{geometry}
\usepackage{scrextend}
\usepackage{booktabs}
\usepackage{graphicx}
\usepackage{multirow}
\usepackage{epstopdf}
\usepackage{float}
\usepackage{tikz}
\usepackage{pgfplots}
\usepackage{subfig}

\DeclareGraphicsExtensions{.pdf,.eps,.png,.jpg,.mps}
\usepackage{amssymb}
\usepackage{lineno}
\journal{TBA}

\begin{document}

\begin{frontmatter}


\title{Uniqueness of Medical Data Mining:\\ How the new technologies and data they generate are transforming medicine}



\author[vcu,pan]{Krzysztof J. Cios\corref{cor1}}
\ead{kcios@vcu.edu}

\author[vcu]{Bartosz Krawczyk}

\author[osu]{Jacquelyne Cios }

\author[har]{Kevin J. Staley}

\address[vcu]{Department of Computer Science, Virginia Commonwealth University, Richmond, VA 23284, USA}

\address[pan]{Polish Academy of Science}

\address[osu]{Ohio State University}

\address[har]{Harvard Medical School, Boston,USA}

\cortext[cor1]{Corresponding author}

\begin{abstract}
\end{abstract}


\end{frontmatter}

\section{Introduction}
\label{sec:int}

Medicine is becoming increasingly dependent on new technologies and they have big impact on its practice. An important aspect of these technologies is that most of them also generate large amounts of digital data that need to be stored. Unfortunately, data, even big data, have no value until they are analyzed with the goal of discovering new and actionable knowledge, which would potentially benefit patients and communities. 

Medicine is becoming Predictive, Preventive, Personalized and Participatory, a rubric known as P4 medicine \cite{Flores:2013}. Using systems medicine, defined as application of systems biology to human disease \cite{Hood:2012}, we are able to predict the probability of a disease. Predictability, in turn, makes prevention possible, for example the implementation of measures to either prevent the disease or reduce its impact on patients. These approaches allow for increased personalization, also known as precision medicine. This realization came from simple observation that drugs, treatments and procedures do not work in the same way in different patients. However, by obtaining additional data about a patient, like genetic profile, the treatments can be fined-tuned. New technologies also support growth of participatory medicine, also known as patient-centric medicine. With the spread of the Internet and social networks, patients and families have moved from passive recipients of healthcare to well-informed partners, who co-decide the course of their treatments.        
     
\textbf{Healthcare data security.} Huge volume of healthcare data collected on patients has potentially enormous value but keeping that data secure poses a formidable challenge. There is a growing importance of data security \cite{HamidRHAA17}, as mandated by regulations such as Health Insurance Portability and Accountability Act (1996) and the Genetic Information Nondiscrimination Act (2008).  The two acts set standards for the electronic exchange, privacy and security of health information, or the misuse of genetic information, e.g., for health insurance or employment decisions. However, no healthcare data security system is foolproof in spite of using the new technologies that better secure their storage and transfers \cite{ChiouL18}. Following the rules that exist in some countries, but not globally \cite{Williams07}, is not sufficient as there is always a human factor involved in securing the data \cite{SusiloW07}. The factors threatening data security range from innocent human mistake that can make data vulnerable to malicious players, to the human greed that opens data to the highest bidder. 

The collection of healthcare data occurs also at social media giant tech corporations \cite{Fernandez-Luque18}.  Data collected on all of us, such as by Facebook, can be and are used for dubious purposes like social engineering and marketing, along with potentially good purposes \cite{DiezCZC16}. Unfortunately we have witnessed huge data breaches (Facebook, Yahoo) that make data available to bad players. Some new technologies can alleviate this problem, for instance, using blockchain technology leaves a trace of every transaction performed, such as sending data from one hospital to another, but this works only for official data transfers, as blockchain cannot prevent data from being hacked or stolen. Because of the latter, we shall talk about \textit{ P5 medicine}, by adding the Privacy-preserving attribute to the predictive, preventive, personalized, and participatory. It is a new task, to develop technologies to better protect health data in the future. 

\textbf{New technologies and medical data mining.} We describe below how new technologies, several of which came into being since publication of our original paper on the uniqueness of medical data mining \cite{CiosM02}, that collect large-scale heterogeneous medical data on individuals and populations, are changing the field of medicine. Examples of data generated by those technologies are image and –omic (genomic, proteomic, etc.) data.  Below, we highlight the biggest changes that have occurred since the Cios and Moore paper \cite{CiosM02}. 

Human medical data are at the same time the most rewarding and difficult of all biological data to analyze \cite{SiulyZ16}. Humans are the most studied species on earth and data and observations collected on them are unique and cannot easily be gained from animal studies. Examples are visual, cognitive and perceptive data, such as those relating to discomfort, pain and hallucinations. Animal studies, being shorter, cannot track long-term disease processes, such as atherosclerosis. Since the majority of humans have had at least some of their medical information collected in digital form, this translates into big data. Unfortunately, analyzing human data is not straightforward because of the ethical, legal, social and other constraints that limit their use. Hospitals are the ones who store majority of medical data, however, many of them have little interest in sharing them, which adversely impacts global healthcare by restricting large-scale data analysis and new findings. While the current trend is towards open-access data and collaborative environments, issues related to medical data privacy and market competition are not easily alleviated.

On the technology side we have seen creation of disruptive technologies such as blockchain, cloud computing, wearable devices, and augmented reality. A disruptive technology does not have to be an entirely new one: often it has been long-existing but was greatly improved in terms of, for instance, speed or accuracy. Examples of the latter technologies are artificial intelligence (AI) and machine learning \cite{Alinejad-RoknyS18}, natural language processing (NLP) \cite{WangLHXSJLDWZZZ15}, image understanding \cite{Demner-FushmanAKM15} and robotics \cite{DaneshmandBBA17}.  They all call for new ways of looking at medical data. As an example, a system called nEmesis, based on NLP, was developed to identify restaurants reported by customers on a Twitter as a source of the food-borne illnesses \cite{SadilekKDLPTS16}. The system collects that and allows health officials to quickly localize impact of a spreading disease. Another example is a system to identify patients who truly need a massive blood transfusion, avoiding frequent overtransfusion, a bad use of a limited resource such as blood bank \cite{Keffe:2008}.     

There is a growing need for medical standardization and benchmarking to allow for fair comparisons of procedures across physicians and hospitals \cite{Staley1}. We illustrate this need and how it was solved in the field of data mining. Before standardization, each data mining researcher had to create their own software to analyze data. This was an enormous waste of time, prone to significant variations among different implementations of the same algorithms. A fair and unbiased comparison of the methods was practically impossible, which reduced their impact and spread of practical applications. To address this issue, researchers developed well-tested software packages for almost all types of machine learning algorithms: examples are MATLAB (commercial) \cite{Kasai17} or WEKA (free) \cite{BouckaertFHHPRW10}. Similar process took place in statistics for validating models generated by machine learning algorithms \cite{BenavoliCDZ17,Demsar06}. These statistical methods also became part of software packages, including in the two just mentioned. Statisticians went even further and addressed problems  with already published papers \cite{CoraniBDMZ17}. A web-based algorithm called statcheck checks validity of statistical testing used in papers; it reads both pdf and HTML formats of papers and recalculates p-values and other measurements \cite{Nuijten2016}. Another statistical example is G*Power software for computing statistical power for $t$-tests, $F$-tests, $z$-tests used for design of experiments to determine effect sizes \cite{Faul:2007}. Developments such as those contribute toward improving transparency and reproducibility of research results, which remains a problem even in top journals, although some publishers addressed this problem by requiring that both data and software must be published along with the reported results.   

There are no similar benchmarks for medical evaluation, management and procedures because medical practice is still a mixture of art and science \cite{VanschorenRBT13}. For instance, is it possible to objectively compare the outcome of two brain surgeries for the same type of cancer diagnosed at the same stage and located at about the same brain region?  Can certain clinical scenarios be similar enough to be analyzed similarly for the good of patients? More importantly, can health care outcomes be fully standardized \cite{WongWW17}?  On the positive side, the metric of patient satisfaction is now tracked nationally since lawmakers have tied this to the Center for Medicare and Medicaid (CMS) reimbursements.  Epidemiologists have been lumping together similar cases to measure outcomes in large population-based studies and some clinical practice guidelines have resulted from them \cite{Babyar18,KorstASF11}. For example, the Framingham heart study \cite{Mahmood2014TheFH} regarding risk factors of cardiovascular disease has given rise to practical management tools for the clinician, such as calculators for the risk of atrial fibrillation or stroke under various conditions. These help identify and standardize conditions under which physicians offer patients certain treatments like blood thinners.  However, more performance-based metrics are still need to be developed. 

\textbf{New technologies demand new way of educating medical doctors.} Because modern medicine heavily depends on technology there is a need for physicians to understand the technologies they use \cite{Sejdic15}. Some physicians deal with this problem by updating their technological know-how on their own, but this is not easy and they can learn only about a few such developments and at only a rather superficial level. As a side note, there are about 34,000 journals and they publish about 3 million papers each year.  

To address the problem of changing technologies, and medicine becoming more “technical”, new ways of teaching it are being implemented to train physicians.  Almost all medical curricula incorporate training in the use of electronic health records and searching for necessary data in databases such as Cerner or Epic.  A drastically different paradigm to medical education, however, was developed and implemented at Carle Illinois College of Medicine, jointly funded by the University of Illinois at Urbana Champaign and the Carle Health System. The College requires that all applicants, in addition to the typical medical curriculum also have knowledge of computer science, mathematics and statistics. Importantly, it is the first school that teaches each medical subject from three perspectives: biological, clinical and engineering. It is done by having three instructors to teach each topic; for instance, while learning about the cardiovascular system they also learn, from different instructors, about fluid dynamics and genetics. During clinical rotations they are required to identify technological solutions that are or can be applicable to the disease/problem investigated. One of the College’s goals is that by utilizing more technological solutions, including software such as AI and machine learning, the cost of medicine will decrease. The latter is especially important in the U.S., which has by far the highest costs of medical practice of any country \cite{Strickland:2018}.

The rest of this paper is organized as follows. Section 2 discusses the issue of heterogeneity of medical data, Section 3 focuses on data mining issues, Section 4 touches upon legal, ethical and social issues, while Section 5 emphasizes special status of medicine. We end with concluding remarks.

\section{Heterogeneity of medical data}

Cios and Moore \cite{CiosM02} in the section under the same heading focused on volume and complexity of medical data, importance of physician’s interpretation, sensitivity and specificity analysis and poor mathematical characterization. All these topics but the first are as valid now as they were then. The volume and complexity of the medical data, however, has increased exponentially as massive amounts of data are being collected daily on patients, populations, as well as on healthy individuals. 

\subsection{Sources of medical data}

The reality is that now it is much easier to collect and store big data than to analyze them. As a result large amounts of the already collected are never analyzed \cite{AlonsoDRHL17}. One of the reasons for the latter is that data capturing  technologies change rapidly and it makes little sense to analyze data collected with the use of old technology. For example, every few years the resolution of images improves by orders of magnitude so analyzing data collected with a much lower resolution makes no sense. At the extreme, it is now possible to use bioluminescence imaging to visualize, measure tumor growth, and even observe cell-to-cell interactions \cite{Iwano935}.   

Out of the ten biggest companies in the world, seven are information technology companies, of which five are American. Although only one of them is a healthcare company (Johnson\&Johnson), almost all of them have large presence in the area of health care. The result is rapid growth of sources and amounts of health data generated. Below we list some sources responsible for medical data exponential growth:

\noindent $\bullet$ EHR (Electronic Health Records). In 2002 less than $20\% $of American hospitals have had fully implemented EHR systems but now more than $90\%$ do \cite{CyganekGKKPWW16}. US hospitals now generate about 50 Petabytes ($50 x 10^{15}$ bytes) of data per year; yet only about $3\%$ of it is fully analyzed \cite{cisco:2011}.

\noindent $\bullet$ Patient rights. Because of increasing patient rights, almost all hospitals now allow patients to see and download their records \cite{IshikawaKTTKIT04}.  Note, that this increases risk of security breaches.

\noindent $\bullet$ Genomic and proteomic data. The cost of analyzing a human genome went down from millions of dollars since its decoding in 2001 to hundreds of dollars now. This translates into exponential increase of genomic, protein, DNA, RNA, nucleotide sequences, protein structure, microarray etc. data collected.

\noindent $\bullet$ PubMed publications. Medical research results, measured by publications, double every few years. For example, the number of genotyping articles, which identify variations in specific pre-defined single polymorphisms within a gene, grew ten-fold since 2002 to over 1000 now. The number of sequencing articles, looking for variations throughout the entire gene, grew from few to about ten thousand. In addition to the published articles, many publishers now also require publishing data and software used for analyzing them; the purpose being that the results can be repeated and verified by others.

\noindent $\bullet$ FDA-approved tests for at home use. The number of such tests, those that do not do any harm to a human, and that are fairly accurate, doubled since 2002.  

\noindent $\bullet$ Clinical trials. Their number has increased by an order of magnitude since 2002.

\noindent $\bullet$ Smartphones and wearable devices. With the proliferation of smartphones and the non-invasive devices came an increase in the number of downloads of health-related apps \cite{JohnsonHPCJM15}. The majority of the latter are for physical conditioning purposes but also for diseases, treatments, stress reduction, diet, nutrition, etc.  

\noindent $\bullet$ Medical imaging data. Medical imaging has made significant advances over the last two decades \cite{GodinhoCO17}. New devices allow for more precise monitoring of not only static but also of dynamic changes. Therefore, visual data now also include time-related information: such imaging is known as 4D imaging (red/green/blue/time). 

\noindent $\bullet$ Social media text data. The role of social media has exponentially grown in the last two decades.  As people are sharing details about their well-being and seek others in similar situations, social media is another big source of medical data \cite{LimsopathamC16}. For instance, Twitter microblogs with geo-localization are used for tracking the spread of viral diseases. Social media has changed the way we look at individual and population-level health issues. Patients seek medical information online on websites such as WebMD and the websites of large medical centers or disease-specific foundations.  They communicate with each other about symptoms and share personal experiences \cite{KimL15}.  Physicians also use social media, such as professional networking sites like Doximity, Sermo and Doc2Doc. These allow using opinions of a community or understand the standard of common medical opinion or practice \cite{Ventola:2014}.  It is difficult to efficiently store and process these data but they contain invaluable insights to individual and population-level health conditions.  The use of social media creates potential new problems such as dissemination of poor-quality information,  misrepresentation of healthcare providers’ words or actions, patient privacy breaches, and the erosion of professional boundaries between physician and patient. 

\noindent $\bullet$ Electronic surveys data. The ubiquitousness of Internet access allows for significant transformation of user surveys that probe the patient’s attitudes and satisfaction with the healthcare service. By  electronic surveying, one has access to a much bigger and more diverse populations of patients \cite{Rebollo-Monedero18}.  

In 2017, worldwide digital healthcare data were estimated at about 10,000 Petabytes and are projected to reach 25,000 Petabytes by 2020. These huge data amounts require use of special computational methods such as distributed processing, using a MapReduce paradigm implemented on platforms such as Hadoop, Spark, or Storm.

\subsection{Data integration}

Dealing with highly heterogeneous medical data requires that they are integrated to take full advantage of different “views” of a patient to make the most correct diagnosis.  For instance, it may be required to integrate patient’s genetic, image, signal, text, numerical and streaming data. Integration aims to find relationships between different data sources to better diagnose a certain medical condition \cite{CalhounA09}. An example is cancer diagnosis, where fMRI data are often combined with genetic data \cite{LiuCWXAM14}. Furthermore, it is beneficial to combine individual patient’s data with population and social media data to take advantage of all information about a disease. Learning from heterogeneous data, however, is not trivial due to differences between structured and unstructured data, varying data quality, their certainty and veracity levels, etc.  

The concept of integrating heterogeneous data collected on single patients with data collected on a population of patients is shown in Figure 1. It illustrates combining signals like EKG (in red), - omic data (in green), image data like MRI (in blue) and laboratory numerical and text data (in yellow). We can imagine the integrated data as a matrix shown for individual patients. Physicians can look at this integrated data to make a diagnosis. However, if similar data on many patients (a population) are collected and analyzed by data miners, they may discover some patterns that are not recognizable at a single patient level, and this combined information, when presented to a physician, can help them to better and with higher confidence diagnose an individual patient.

\begin{figure}[h!]
	\centering
	\includegraphics[scale=0.75]{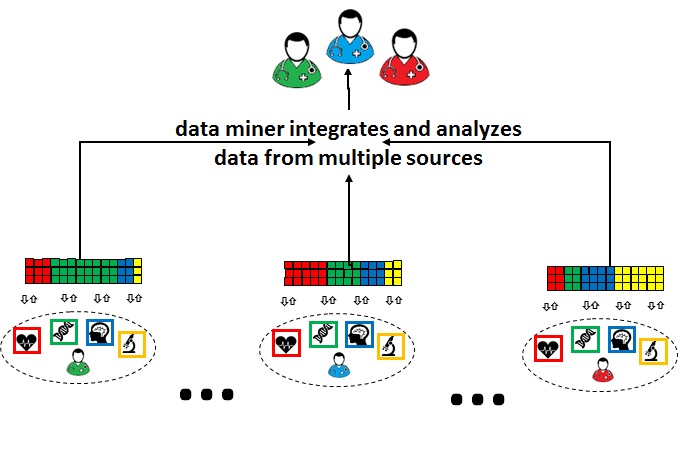}
	\caption{Multi-source view of a patient and the data fusion operation.}
	\label{fig:1}
\end{figure}

In other words, population data allow for discovery of more reliable patterns of diseases than just looking at individual patients. For instance, researchers in Europe analyzed 15,000 patients and found that people actually have five distinct types of diabetes, not just Type 1 and Type 2 \cite{Ahlqvist:2018}. This would have not been possible without the population-wide study and without using accurate clustering algorithms. 

Dealing with both static (e.g., images) and dynamic data (e.g., streaming data) makes analysis of data even more complex, as the dynamic data instances arrive continuously. Examples of streaming data include signals from wearable devices, from in-house sensors, text from social media, or videos capturing movement of patients at home \cite{ZhangYSKM18}. Such data require constant updating of machine learning algorithms \cite{KrawczykMGSW17} and recognizing when statistical characteristics of the incoming data change over time; a phenomenon called concept drift \cite{Ramirez-Gallego17}.  Example is a patient initially diagnosed with a breathing problem but more incoming data may change a diagnosis to a heart problem.   

In an intensive care unit patients are connected to multiple sensors and all their measurements are  observed at some control monitor \cite{SalmanRSS14}.  Doing so allows for detecting disorders more quickly and reliably than from a brief bedside observation, and thus address them without a delay. Such intraoperative monitoring has become standard of care for brain and spinal cord surgeries.  In this scenario real time EEG and evoked potential signals are read by a monitoring physician while a patient is undergoing various brain, carotid artery or spinal cord surgeries. For example, in a carotid endarterectomy, monitoring allows the physician to detect shunting of blood away from a potentially vulnerable region of tissue before irreversible ischemic injury occurs. One of the devices used in ICUs is a mechanical ventilator. One problem with their use is that quite often there exists a dyssynchrony between the patient’s own breathing pattern and the steady one forced by the mechanical ventilator, which causes patients to “fight it”, which may do more harm than good for a patient. However, by measuring airflow and using machine learning algorithms detecting in real time the level of dyssynchrony, an adaptive controller can constantly adjust the mechanical ventilator airflow to keep it synchronized with the patient’s breathing \cite{GholamiPHCMPB18}.

\subsection{Medical Internet of Things (IoT)}

Medical IoT is one of the fast growing fields. The idea behind IoT lies in creating a network where people and their smart devices are interconnected. Smartphones, smartwatches, and smartbands offer on-board sensors as well as significant computational power that allows them to collect, transfer, and even to some extent process the data as they arrive.  As such IoT offers new opportunities for individual- and population-level medicine \cite{RathoreAPWZ16}. By using lightweight wearable non-invasive sensors, continuous streams of data are collected on patients in their home environments, without endangering their privacy \cite{MezghaniEDSP15}. Biomedical signals, activity patterns, movements, and behaviors can be observed and analyzed in near real-time, giving valuable information about patients. On a population-level scale, data coming from thousands of patients are analyzed to find general patterns.Some of IoT characteristics:

\textbf{IoT network architecture.} It is a backbone of IoT \cite{KangSBG15}, responsible for collecting, transmitting, and protecting data, as well as enabling communication among the devices \cite{KohaniP18}. It consists of: 

\noindent $\bullet$ Topology, which represents different components of medical IoT and scenarios \cite{KhanGWGK17}. It consists of data providers (humans, sensors), resource providers (hardware used for computing), and coordination space responsible for handling data transfers, requests, and results. Sometimes, a physician or other healthcare professional, such as an athletic trainer, is involved.

\noindent $\bullet$ Specification, which refers to IoT physical elements, their organization, principles under which they operate, and what specific techniques are used. Popular ones include interoperability to wireless networks (WAN), capability for high-speed data streaming, secure communication between devices, and data back-ups. 

\noindent $\bullet$ Platform, which refers to the used computing and software solutions, such as data centers, high-performance computing clusters \cite{Cano18}, MapReduce approaches \cite{ColosimoPMH11}, and software capable of analyzing data streams in real-time \cite{KalidZZSHH18}, such as SAMOA \cite{MoralesB15}.

\textbf{Healthcare services in IoT.} Medical IoT offers an ever-expanding array of services that are becoming integrated with the society lifestyles; some of them include:

\noindent $\bullet$ Ambient Assisted Living – attractive for elderly and people with disabilities \cite{SaldanaMPV18}. Instead of relying on human supervision, patients are assisted by intelligent systems capable of monitoring their condition, improving their physical and/or mental condition, helping with achieving daily self-care goals, and alerting a health care provider in case of need while preserving patients’ privacy.

\noindent $\bullet$ Community Healthcare – allows to collect spatio-temporal data on communities of varying sizes. This enables analyzing trends and patterns at population-level, finding dependencies between different groups of patients, and analyzing evolution of correlations among multiple health factors. It is used by patient populations to benefit from social support, which social networks bring. In a study \cite{Chaudari:2017}, traumatic brain injury patients were asked to log their self–ratings of emotional or cognitive outcomes on a regular basis and supporters were able to  comment on these scores through a cell phone app.  It was shown that participating patients improved performance scores compared to non-IoT connected controls. 

\noindent $\bullet$ Semantic Medical Access – works with natural language with varied semantics and ontologies \cite{DrozdowiczGP16}. This includes voice command interfaces for smart houses, analyzing unstructured clinical notes, or extracting useful information from bot-driven conversations with patients. 

\noindent $\bullet$ Embedded Context Analysis – requires understanding of the context related to a specific domain, as well as incorporating the domain knowledge into machine learning \cite{UbayashiKHT11}. This is important in particular for dealing with continuously arriving streams of data that need to be understood from analytical and medical perspectives.

\section{Medical Data Mining}

The term data science is now often used instead of the term data mining so they are used here interchangeably.  In the presentation below, we follow the knowledge discovery process (KDP) \cite{CiosM02} that consists of six steps: Understanding the domain, Understanding the data, Data preprocessing, Model building, Model validation, and Model deployment \cite{CiosPS98,Cios:2007:DMK}. The KDP process is highly interactive between data miners and data owners, the latter being patients, hospitals and physicians. It is also highly iterative since if a generated data model is not sufficiently accurate (as shown during the validation step) data miners have to go back to better understand the domain and data, perform different data preprocessing operations, etc. \cite{Schu:2007,ShinSSDWSC09,Staley:2008}. One should remember that the most important and time consuming of the KDP’s six steps is data preprocessing, which takes between 50-70\% of the entire data mining effort.  

Thanks to technological advancements, the KDP process is now partially automated. The reason behind this development is that a data miner is needed to analyze data (and even they spend considerable time doing so), companies designed systems that to a large degree automated the steps of: preprocessing, model building, and model validation. TPOT, is an open-source tool that automatically optimizes feature selection methods and data models to maximize accuracy \cite{link1}. Another system, DataRobot \cite{link2}, tests and refines data models; it operates on structured data but can also find key phrases within unstructured text as well. Quill (Narrative Science) system \cite{link3} is an interesting system that generates from raw input numerical data a data model in a natural language, i.e., in English, using Natural Language Generation to write a “story” about the most important information found in the data; it takes as input structured data only. Loom Systems \cite{link4} take as input structured and unstructured data, either as batch or as streaming data, and visualizes the generated data models. A popular and free machine learning platform already mentioned, WEKA, offers hundreds of algorithms for data preprocessing, clustering, classification, regression, visualization, etc. Its Auto-WEKA version allows for simultaneously selecting a learning algorithm and setting its parameters for selecting the best model \cite{KotthoffTHHL17}. MOA is popular open-source package for analyzing streaming data \cite{BifetHKP10}.

Another known tool is IBM Watson \cite{Ferrucci12} (commercial) that gained attention in healthcare arena as it is capable of answering questions formulated in natural language \cite{LallyPMBPFFC12}. It can model any domain, identifying unique annotation components specific to it. It can process unstructured data such as books, journals, health records, e-mails, and voice messages. It greatly reduces human effort and allows for a semi-automated knowledge discovery. It can also perform evidence-based learning, which allows for creation and improvement of clinical decision support systems. IBM used Watson to improve cancer detection and treatment. IBM says Watson will be capable of storing a knowledge base comparable to over 1000 oncology experts and thus offer a new outlook on cancer data analysis. It should be noted, however, that Watson lacks humanistic data inputs that only physicians can bring to the decision-making equation. For example, it cannot read facial expressions and know when there is something a patient is not comfortable with. It cannot understand preconceptions about patient’s own disease or experiences of friends and family members. It does not take into account changing social situation, with whom and where a patient lives, ability to access healthcare providers, general safety and daily habits, or patient moods. Notwithstanding its weak points Watson, or a similar product, has a potential of becoming a game-changing technology for healthcare.

It should be noted that the way in which physicians collect data, far from simply going through a checklist,  includes getting to know a patient emotionally and psychologically; this process in itself is both diagnostic and therapeutic. Physicians know that just listening and holding the hand of the patient is therapeutic. For these reasons, medicine may always be an art and a science.

\subsection{Medical data preprocessing}

Due to the ever-growing size of medical data and their heterogeneous nature two basic operations are always performed: preprocessing of the data and their integration. The latter was already described above.

Recall that preprocessing is a very time-consuming step. Some preprocessing methods are general, whereas other are applicable to specific kinds of data, e.g., only to signals or images. When we talk about dimensionality of data it in fact refers to three aspects of data: a) the number of data instances, like the number of patients, b) the number of features/attributes describing data instances, like different tests performed on a patient, and c) the number of values each attribute takes on, such weight that is in the range 100 to 500 pounds. The basic preprocessing operations are instance selection, feature selection, and feature discretization. An instance (data point) is a singular case of data, such as patient.  A feature (attribute) is an observed variable describing each instance.  

Instance selection (IS) focuses on choosing the most important instances while discarding other, for instance discarding wrongly diagnosed patients data. It is used to remove outliers created by human or mechanical error during data gathering.  IS is especially important in the case of imbalanced data, where one of the groups is underrepresented compared to the rest, e.g., having many sick patients data but few normal patients data. 

Feature selection (FS) is the process of finding the best feature subset from the original set of features \cite{GuyonE03}. See Figure 3 as an example, where only two features were selected for decision making from the four features measured for a patient. FS is a key technique used with analyzing medical data where it is important to retain original attributes for model building.  Feature extraction (FE) methods, on the other hand, involve transformation of original features into a new feature space using techniques such as principal components analysis\cite{Pearson:1901}, or neural networks \cite{ShinSSSRMKC10,SwierczCSKAS06}.

Another basic preprocessing operation is discretization, which reduces the number of values a feature takes on. For example, age can be in a range from 0 to 110 years but can be discretized by grouping ages into intervals such as infants (0-1 year), children (1-12), teenagers (13-18), young adults (19-30), etc. Discretization is performed with or without taking into account class information (like diagnosis). Some algorithms, like CAIM, use the relationship between the feature being discretized and class to accomplish discretization without the need for a user to specify the number of bins/discrete intervals \cite{KurganC04}.

Majority of data mining methods operate on vectors. While this is sufficient for dealing with many types of medical data, analyzing more complex data structures using vectors can be challenging. Let us consider video sequences, such as a 4D ultrasonograph, which consist of two-dimensional frames with three-valued pixels (in RGB system) that are displayed many times per second, thus adding the 4th feature (time) to data description. In this case, as well as for other images such as CT, MRI, PET-CT, SPECT, and fMRI a tensor representation (basically a matrix) is better than vector representation as it preserves spatio-temporal relationships \cite{Yin18}.

\subsection{Model building}

We do not describe here various types of machine learning algorithms used for model building as they are well covered in many articles and books on data mining \cite{Cios:1998,Cios:2007:DMK}. Instead we comment on deep machine learning, one of the disruptive technologies.  

With advances in image acquisition technologies and their growing use in medicine it became obvious that computers need to be used to perform fast analysis of hundreds or even thousands of images. There are significant variations between different populations of patients, like male vs female, Asian vs African-American, etc. The differences can be subtle and not easily identifiable by the human eye in the images.

Using techniques such as deep learning (DL), especially one called deep neural networks (DNN), such differences can be easily found \cite{LanWFLWD18,LitjensKBSCGLGS17}. DNN can automatically select the most important features and use them to distinguish between \cite{Cios18}, say male and female CT images of the heart.  The use of DL in a medical setting may be challenging due to the fact that it usually requires large data sets to achieve good accuracy, and such data are not easy to acquire for each disease type. Fortunately, new algorithms can be trained on much smaller data sets \cite{Georgeeaag2612}. Requirement for having a big dataset can be alleviated by the usage of the technique of data augmentation, which increases the volume of the underrepresented images by adding additional images that are distortions to the original images, e.g., distorted by changing brightness or adding noise. 

In spite of great successes of deep neural networks, researchers have also shown their spectacular failings on image recognitions tasks, which are however trivial to humans.  For instance, a trained DNN was tested on only slightly modified original images used in training. The modification was such that that there was no perceptible difference to the human eye between the two. Unfortunately, when the modified original image was used for testing, the network failed to recognize it \cite{SzegedyZSBEGF13}.  In another work, the researchers modified the image used in training in such a way that it had no resemblance whatsoever to the original image to the human eye. When used for testing, this image appearing to the human eye like a TV static noise was not only recognized as a peacock (the original image) but also with a high accuracy of 99.6\% \cite{NguyenYC15}.

Creating an efficient data analysis algorithm is a challenging task.  A single “best” algorithm does not exist; the performance of any algorithm often depends on the type of data being analyzed. This requires careful selection of a specific algorithm and its parameters, for a specific task. Furthermore, as we gather more data we cannot assume that the previously trained model is still valid. This requires involvement of a data mining expert and not blind use of any machine learning software package, which will always generate some "results". Data mining experts, however, may not be available in smaller hospitals.

\subsection{Validating generated models}

Validating the generated models of data consist of two stages. First, a data miner uses formal techniques to assess their quality, such as its goodness of fit (accuracy) and predictive power.  Second, the model (information/knowledge) is presented to medical domain experts who have the final say on whether what was discovered is truly novel and useful. Only the best model, as assessed by domain experts can be used in practice. 

\textbf{Formal techniques}. They include methods analyzing how well does a given model performs on training and test data. The approach is to use metrics such as specificity and sensitivity, geometric mean, F-measure, Area under the ROC curve, Kappa statistic etc \cite{DBLP:books/cu/Japkowicz2011}. Apart from performance, a data miner also evaluates stability of the model, i.e., how the results change with small changes in the input data. In cases where the speed of building a model is important, one should use simpler methods before using possibly better but more complex ones. Finally, statistical tests are used to compare different models to check if the differences between them are statistically significantly different, examples are ranking (e.g., Friedman test) and post-hoc (e.g., Shafer test).

\textbf{Expert evaluation}.  Even if a model was found using formal techniques as good and statistically better than other models, a medical expert must evaluate it and be able to understand why the model is suggesting a specific diagnosis. Remember that the quality of the generated model always depends on the quality of the training data. We also must account for imperfections in the data; a machine learning algorithm may find some correlation in data that would be dismissed by a physician as wrong, e.g., a correlation between flu resistance and income (which may be true but not valid).

\subsection{Deployment}

Once a model has been validated both formally and by medical experts, the last step in the KDP is the model’s deployment. We describe their deployments by means of case studies. They illustrate how  advancements in machine learning, imaging, AI, and other technologies change medical practice.       
  
\noindent Case study 1: How AI and high resolution imaging changed telemedicine

Telemedicine, aka telehealth, is the use of information and telecommunication technologies to provide health care service from a distance without a personal visit by a patient. It was first used by the US military when radiology images were sent over telephone lines. Universities then used a two-way television system to transmit medical information across campus (University of Nebraska) or to transmit electrocardiography signals from fire departments to speed up diagnosing patients in emergency situations (University of Miami). In the 1960s NASA, using microwave technology, provided health care services to Native Americans as well as to orbiting astronauts. The spread of telemedicine, however, reached a plateau around 2000s because its widespread use required sending high definition images and videos through broadband infrastructure, which was lacking in many communities. New or greatly improved technologies along with better broadband coverage led to its current massive-scale use. The two technologies that played a key role are machine learning/artificial intelligence on the one hand, and high-resolution imaging on the other hand.  The reason for the latter is that imaging devices “see” (their resolution) on the orders of magnitude better than humans; the same is true for hearing devices. Because of these developments it was possible to leverage the power of hardware, accompanied by software controlling it, to augment human perception. These technologies allow for realizing the collaborative human-computer systems that take advantage of strengths of machines and humans, complementing each other. 

AI research started about half-century ago, with great promises, but only recently became sufficiently fast and accurate to be used in real time. This progress was due to high-performance distributed computing systems, like Apache Spark, and smart algorithms, like for deep learning.  As a result, telemedicine, because of these two transformative technologies, started to be used on a massive scale. One case in point is Mercy Health System in Missouri that built and pioneered the first hospital dedicated entirely to remote care of patients, called Mercy Virtual (MV). It serves almost a million patients in the Midwest. The MV hospital, without beds but staffed with nurses and physicians, uses state of the art information technology but patients can use just their smartphones or tablets to collect and transfer much of the data requested by a physician, or go to a local clinic to undergo more specialized tests that are transferred to MV. Importantly, before a physician even sees patient data, they are first processed by AI techniques, resulting in triaging patients into groups so that physicians can attend to the most seriously ill first. Using AI also speeds up the process of diagnosing, as it automatically suggests “diagnoses,” with a low misdiagnosis rate and avoiding human fatigue which is a significant contributor to medical errors.  Advantages of telemedicine include savings on waiting and travel time for patients, quick access to a physician in case of emergency, and low-cost monitoring of permanent health conditions, such as diabetes. 

It is known that diabetes at later stages of the disease causes diabetic retinopathy, which eventually leads to blindness. In April 2018, the US Food and Drug Administration approved the first-ever AI-based software (developed by IDx LLC) for screening diabetic patients for early detection of the disease. The inputs to the software are images of the retina taken by high-resolution digital cameras. Many researchers were working for a long time on automating retinopathy quantification and diagnosis \cite{CiosSS02}, but it takes a company to bring the ideas to the market. The IDx software was almost 90\% accurate in tests for both identifying (test positive) the diabetic retinopathy or excluding (test negative) it.  Only the patients who tested positively would see an ophthalmologist. This cuts down on wait times to see an ophthalmologist and use of society’s resources, e.g., Medicare/Medicaid dollars.  It may seem that accuracy below 90\% is not good enough but human experts do not perform better using their (low-resolution) eyes; this is another example of the growing human-computer complementation of skills.      

We envision that telemedicine services will grow significantly in the near future because of new technologies, including those mentioned above and the new ones that will emerge in the future.

\noindent Case study 2: How AI is changing medical diagnostics

China is one of the first countries to use AI technologies on a larger scale to automate the process of  medical diagnosis. It has a great need for such automation because of the size of its population and a small number of physicians per capita, as compared with the other countries. Interestingly, Chinese agency, equivalent of the FDA, determined, in 2017, that AI-based diagnosis is a medical “device” and as such needs to be approved by it before it can be used in medical practice. Building on great successes of deep neural network algorithms in image recognition, the most obvious medical areas for developing first automatic diagnostic algorithms have been in the areas of radiology and pathology. In fact, this process has been, at least partially, already automated in several countries, including in the US and India. An algorithm was developed in China to read/diagnose lung images. The major difference with the situation in China as compared with the US is that the software is already used at several Chinese hospitals, which deal with thousands of patients per day. Other Chinese groups develop programs for designing dentures, or for detecting formation of blood clots in lymphoma treatments \cite{link5}. Time will tell whether China will become a leader in automated medical diagnosis. The advantages that China has over other countries, however, are big and many. For one, it has less stringent rules about patient data governance than those in place in the EU or the US. The other one is the top-down way the country is governed so once it has been decided that medical diagnosis automation is a priority, sufficient funding for achieving this goal is made available and their companies and scientists have a strong incentive to work on it. Additional strong motivation for the scientists is knowledge that, once developed, their “products” will be used in day-to-day medical practice.  

All scientists hope that their work will result in something that can be used for a common good. And what is more rewarding that knowing that one contributes to human well-being?

\noindent Case Study 3: How AI is changing medical diagnostics

AI has potential application in a decision-making process in the field of epilepsy, where 1/3 of patients are medically refractory, meaning that they cannot become seizure free despite the use of at least 2-3 medication trials at high dosages.  The recurrence of seizures has tremendous social and cognitive consequences, including social handicaps and difficulties with memory, and communication and alertness.   In those refractory cases, the greatest chance of seizure freedom is surgical resection of the brain tissue which represents the origin of the seizure; applicable only to those cases in which there is a single, identifiable seizure origin and is far from salient areas of brain which serve important cognitive or motor functions \cite{Pitkanen01102015}.  In epilepsy surgery candidacy evaluation, to find the seizure focus physicians employ multiple data points from MRI and/ or SISCOM images, co-registered MRI with ictal SPECT put together with seizure localization using EEG, neuropsychological profile from batteries of cognitive and personality testing, interictal PET scan, and fMRI. 
 
Frequently these data are discordant. For example, the scalp EEG electrodes identify the right temporal lobe as the seizure focus, but there is a finding of uncertain significance in the left temporal lobe on MRI and the ictal SPECT shows uptake in the left frontal and temporal regions.  Therefore, there is uncertainty about the focal area of seizure onset.  Intracranial EEG recording is then performed to increase the diagnostic sensitivity and specificity of EEG.  The probability of ascertaining the accurate and precise seizure focus differs depending on a number of predictors including electrographic features of EEG, seizure phenomenology, the lobe suspected of harboring the seizure origin, etc. The best outcome, seizure freedom achieved by surgical resection, is related to the chance of accurately pinpointing the seizure origin \cite{Mcintosh:1993}.  

There are at least two ways in which AI can help. First, use of algorithms for analyzing imaging such as MRI in a more objective manner than analysis by the human eye can help improve the diagnostic sensitivity and specificity of MRI findings. For example, in adult onset refractory epilepsy a common MRI finding is mesial temporal sclerosis, defined by hippocampal atrophy and MRI T2-weighted hippocampal hyperintensity. With the discovery that abnormal hippocampi also have an MRI imaging property called abnormal T2 relaxation times, a data analysis called quantitative magnetic resonance T2 relaxometry has been used to detect evidence of a unilateral abnormal hippocampus \cite{Jackson:1993}.  In addition, software using volumetric measurement is available to characterize brain structures in a more objective and accurate way than the human eye: FreeSurfer and Neuroquant are FDA approved examples.  Neuroquant is in wide clinical use to measure temporal lobe and hippocampal volumes, which improve diagnostic accuracy of entities such as hippocampal sclerosis. Good results have been reported for mild traumatic brain injury and Alzheimer’s dementia population data \cite{ZhangSY17}, on 21 brain regions evaluated \cite{Ochs:2015}.

In the future, AI/machine learning will be used to more accurately put together all of the data points, from both structural and functional brain studies, in patients undergoing epilepsy surgical candidacy, and potentially others such as how long the patient has had epilepsy, the presumed etiology, the family history, etc., then a more accurate analysis can be presented to the patient of how good their prospect of having seizure freedom is with the resection procedure.  This would alter the face of clinical practice and the standard of care, personalizing medicine to a greater degree for those patients.

\noindent Case study 4: How smartphones, wearable devices and social networks impact population health

The issue of elderly care is becoming more important and costly for many countries with aging populations. The fact is that more seniors want to maintain high quality of life, however, offering constant human-based monitoring for them would be prohibitive. A global trend of people living longer brings the need for allowing them to age in their own homes, while still offering medical assistance that is traditionally attributed to hospitals. Fortunately, the medical IoT technology allows for continuous remote monitoring of patients in their homes while being non-intrusive \cite{KINGSLEY2019677}. Below we describe a few such solutions. 

Canary Care Company deploys battery-powered wireless sensors instead of using wearable devices. The sensors are placed in various locations through the home to constantly monitor the senior’s activity. Their system uses mobile data transfer, so any caregiver, such as family member, has access to a secure online account allowing them to check the status of the patient. TruSense is a system that creates an in-home network of mobile devices. These include motion sensors, contact sensors, smart outlets, hubs, speakers and text-to-voice devices that provide a two-way communication between a senior and a health care provider or a family member. GPS SmartSole is another product that combines advantages of mobile devices with inconspicuous design. It embeds sensors into shoes that an elderly person wears, fusing information from GPS, cellular, Bluetooth, and Wi-Fi to monitor the state and location of a person. This offers a unique advantage for patients suffering from Alzheimer’s, as the product is capable of finding a patient that has wandered off \cite{AlberdiWSCABB18}.  One of the largest, the STANLEY Healthcare systems are used by over 5,000 acute care hospitals and 12,000 long-term care organizations or relatives. Their real-time location system is connected with each patient’s electronic medical record. Therefore, a hospital may access real-time data regarding the location, state, and history of each patient under their care \cite{HwangboYJHJ13}.

\noindent Case study 5: How big data and data mining techniques allow to model and predict epidemic outbreaks

Spreading of an epidemic can be modeled using mathematical models fitted to the historic data by analyzing spatio-temporal information gathered over multiple outbreaks in multiple places.  Having such a model allows for predicting their potential outbreaks and how will it behave over time. This, in turn, allows for better preparation for large-scale disease outbreaks, as well as organizing preventive measures among a population \cite{BouzillePCACCLC18}. An example of using big data to improve understanding of viral diseases was demonstrated by IBM targeting a dangerous and difficult to contain Ebola virus. The 2014 West Africa outbreak caused deaths of more than 11,000 people, with more than 28,000 cases reported in Sierra Leone, Liberia, and Guinea. Ebola virus can be transmitted to a human via direct contact with a host animal (most commonly snake or a bat), thus entering the local human population. Then it spreads causing outbreaks that are challenging to contain.  Analyzing data from previous Ebola outbreaks lead to identification of new animal reservoirs, thus increasing population awareness in high-risk regions \cite{AyeniMO15}.  IBM used adaptive models of disease dynamics to accommodate for such properties as virus spreading through direct contact with infected blood or body fluids (i.e., urine, saliva, feces, vomit, and semen), contaminated objects (e.g., needles) and infected animals. They highlight the importance of the fact that men who have recovered from the disease can still transmit the virus through their semen for up to 7 weeks \cite{PerscheidBHJMLA18}. 

U.S. Centers for Disease Control (CDC) also used big data techniques for the Ebola surveillance in several African countries based on collecting and analyzing in real-time text reports coming from mobile devices. These cheap and fast sources of information give healthcare professionals an edge, allowing them to track in real-time the evolution of the disease outbreak and spread.

\section{Ethical, legal and social issues }

Medical data mining must operate under specific conditions and rules, which distinguishes it from other applications of data mining. One is a close interplay between data mining experts and medical experts, but there are other factors to be considered when addressing a medical problem with the help of data science methods.

\subsection{Privacy-preserving medical data mining}

A key element involved in the quest for preserving data privacy while allowing access to the data to researchers is de-identification (aka anonymization) of patient records \cite{FaravelonV10,KouPSC07}. The NIH classifies research involving humans as Human Subjects Research (HSR). Privacy threats come from the possibility of deducing direct identifiers, quasi-identifiers, and other attributes from the data which can be subject of identity, membership, or attribute disclosures. Privacy models, such as k-Anonymity, allow assessing privacy risks of the already anonymized data.  Anonymization can also be done by statistical analysis, i.e., removing high entropy variables. In addition to structured information, medical data contain unstructured text, from which identity information can be removed only to a smaller degree by using natural language processing (NLP) tools \cite{AngFTTSFLHS16}. One application uses an open source solution for statistical identity scrubbing with little human effort.  Disassociation is another method impeding identification of a patient. It is achieved by partitioning data into several pieces and thus allowing for independent processing of the pieces. When dealing with image data, the anonymization also requires checking that no identity information is inserted into images. One study showed that anonymous sharing and cooperative processing of clinical and signal data via web services on a multi-center study on deep brain stimulation for Parkinson's disease was successful \cite{PereiraPWHAP19}. 

Privacy issues in medical data analytics are directly related to distributed data mining without releasing any information regarding the nature of data stored in any of the sources; this is known as Privacy-preserving data mining \cite{ZhangCZCH18}. 

The European Union’s (EU) General Data Protection Regulation (GDPR) \cite{link6} of 2018 regulates the use of personal data, namely, any information relating directly or indirectly to identifiable people. It is the most radical change in data privacy regulations, which affects collection of data on EU citizens, their handling and analyses by companies that operate in the EU marketplace. Data, in particular big data, as such have no value until they are analyzed. Because of the GDRP, however, certain restrictions also apply to AI and ML algorithms used to analyze human data. Humans, however, have several rights, for example, article 20 states “the data subject (read: patient) shall have the right to receive the personal data concerning him or her, which he or she has provided to a controller (read: person/institution storing data)” as well as the right to data portability. Article 17 specifies the “right to be forgotten.” GDRP also strives to allow for easier use of human data for scientific and medical purposes: Article 6(1) states “data subjects should be allowed to give their consent to certain areas of scientific research when in keeping with recognized ethical standards for scientific research.” 

The US and EU have agreement on data sharing, known as the Privacy Shield, but the rules governing human data use are weaker in the US (e.g., Facebook data breach) than in the EU.   
Of more interest for us here is Article 15.1 that states that the data subject has the right to obtain from the controllers (people responsible for data management) information about the operation of algorithms used for analyzing data. Although some machine learning algorithms are easy to understand, such as methods that use original features and generate models in terms of "IF some features and their values THEN a certain disease" rules, many are "black boxes", such as artificial neural networks and deep learning algorithms, that lack transparency leading to little, if any, understanding of how they arrived at their “decisions.” Thus how can one be sure that a decision/diagnosis is accurate? GDPR talks about the "right to explanation" and the right to obtaining “meaningful information about the logic involved” in machine learning, ensuring accountability of automated decision-making. This could be interpreted in various ways, one of which might be requiring explanation of diagnoses suggested by algorithms operating on human data.  This could range from simply providing basic explanation of how an algorithm works, including specification of its parameters and their values. But it could also mean providing all the data on which the algorithm was trained and tested, plus actual inputs (patient data) used in arriving with a certain diagnosis.  

Moreover, since ML algorithms almost never achieve 100\% accuracy, which depends on the size of the training data and the (strong) assumption that new data for which the diagnosis is being made are “similar,” meaning that the new data come from the same distribution as the training data. This is especially important for healthcare data where not only statistical properties of data but also the background knowledge regarding the analyzed disease must be incorporated and used. For example, if a model was trained on Asian patient data for a specific disease it should not be used for prediction on African patients because the prediction results may be entirely wrong, given the genetic and other differences between the two groups. On the other hand, Asians and Africans may be very “similar” in many regards; for example, if both populations have a relatively high prevalence of a severe variant of multiple sclerosis, then they may have genetic similarity at salient loci.  In general, we stress the need of a physician involvement in semi-automated diagnosis process.

Even if the machine learning model is well understood and used on the same patient population, it is difficult to say with certainty that a prediction/diagnosis made by it is correct. Typically, if the model has “seen” during its training data similar to the new case, then the result is probably correct. But how is the physician to know that? Fortunately there are some techniques that help in addressing this problem by providing, in addition to diagnosis, a confidence in the generated prediction \cite{parzen1962}.      

For example, researchers using Mt Sinai’s EHR data on 700,000 patients developed a deep neural network model, called Deep Patient, which was able to predict onset of 78 diseases \cite{Riccardo:2016}. The system performed better than other systems, in terms of accuracy of predictions, for whether patients will get sick based on their data. The system performed well for predicting onset of diseases such as diabetes, schizophrenia and some cancer types, for the time frames of 30, 60, 90 and 180 days, but poorly for some other diseases. Deep Patient can be used as an aid by physicians to alert patients that they are at risk for getting a particular disease. The problem, however, is that deep neural networks are not interpretable. In other words, Deep Patient would not inform physicians why it has made a particular disease prediction for a given patient, although the prediction is quite reliable because it was based on analyzing data on hundreds of thousands of patients’ data.

\subsection{Explainable machine learning}

There is an ongoing research for developing explainable machine learning/AI techniques, aka XAI, because of the problem with understanding of models generated by them \cite{MittelstadtF17}. In addition, the models cannot communicate with a human, and most often are not able to provide answer to a question "WHY this diagnosis was suggested?".  DARPA was the first to realize the need for explainability of algorithms used in decision making, and found that different algorithms had different levels of understanding, as illustrated in Figure 2.  As can be seen, the easiest for interpretation, in terms of explainability, are decision trees. The reason is that their results are written in terms of rules of the form: IF symptom1=x and symptom2=y and … THEN Diagnosis Q; they, however, are accompanied by low prediction accuracy.  At the other end of the spectrum is deep machine learning with high prediction accuracy but very low explainability \cite{ChooL18}.

\begin{figure}[h!]
	\centering
	\includegraphics[scale=0.55]{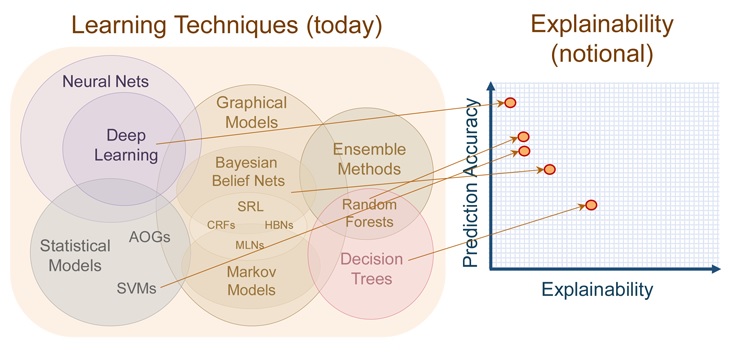}
	\caption{Learning techniques and their explainability (source DARPA (Defense Advanced Research Projects Agency)).}
	\label{fig:2}
\end{figure}

An example in a medical setting is shown in Figure 3. For a new patient data, described by four symptoms (as listed there), a previously trained model is used to do the prediction of a flu.  If the model is explainable the physician can ask a question "why did you decide on the flu?" and the model would "answer": because two symptoms were important in making my decision: "migraine was strong" and "temperature was high".  Seeing this explanation, i.e., reasoning of the system, the physician would be comfortable in accepting the suggested diagnosis. Using such automated programs for suggesting diagnoses, would lead to decreasing  thousands of diagnostic errors made in the U.S. hospitals each year (up to 80,000 according to some estimates). 

\begin{figure}[h!]
	\centering
	\includegraphics[scale=0.45]{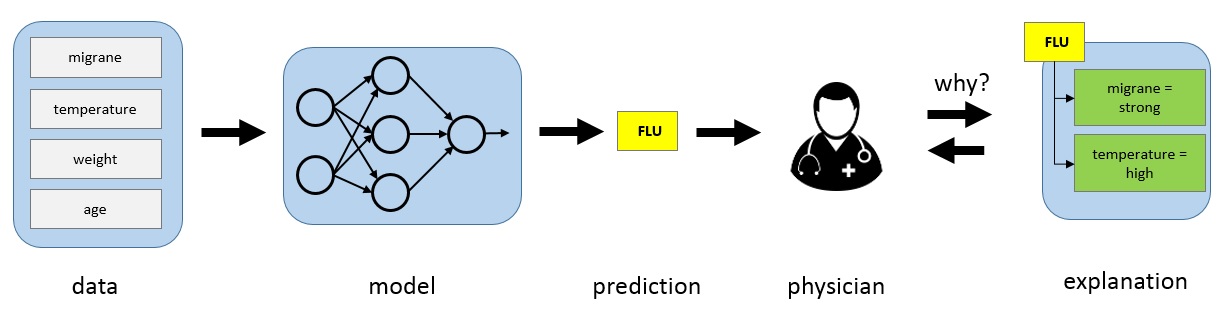}
	\caption{Explainable machine learning example.}
	\label{fig:3}
\end{figure}

\subsection{The ownership of data}

One of the challenges of medical data mining is the problem of data ownership that is related to the issue of Privacy-preserving data mining. Patient data belong to a patient who shares it with his/her physician. But what happens if data was gathered using federally-sponsored research? Does a patient still holds full control over data? The issue of anonymization comes into play but what happens when data was already anonymized? To whom does it then belong?  To a patient, even if he/she cannot link back to her original data? Or to a physician, a data miner, a hospital, the principal investigator of a grant, or the funding agency \cite{Murray2011}? 

These issues are further complicated by a growing pressure from the public and funding agencies that all research must be reproducible, which requires open access to data and algorithms used in achieving published results. On one hand, this allows for a full reproducibility as anyone could access the collected medical data. On the other hand, this calls for better methods for data anonymization \cite{KooH18}. At some point of open data platform development, there will be no distinction between a scientist who collected data (the owner) and a scientist who just downloaded the data for analysis. This is a double-edge situation, as data should be shared to allow for medical advancements, but it can discourage researchers who spend lots of time and effort to generate and curate medical data.

\subsection{The lawsuit issue}

Medical diagnosis that heavily depends on data mining/AI processing of data will sooner or later be subject to lawsuits. The problem is confounded by a series of steps leading to a potentially wrong diagnosis \cite{WillisS17}. Is it always a physician, who relied on the semi-automated clinical decision support system who is ultimately responsible? Or a data scientist who designed an algorithm, software engineer who implemented it, a nurse and/or technician who collected the data? Can the fault be proven at all in some cases?

We need to remember that AI/machine learning systems also can and do make mistakes, in particular when trained on small data sets. Software being a “product” will be subject to liability law, even if it cannot be determined why the algorithm failed to recognize a disease correctly. There is an ongoing discussion how to address this issue, especially in case of personal injury and other losses caused by the use of machine learning \cite{KaranasiouP17}.  There are three scenarios for use of machine learning systems. The first one is when mentally dependent person commits an offence. In this case, a person is deemed as innocent, but anyone who instructed him/her may be held responsible. In case of machine learning, a programmer or user may be held responsible. Second scenario is a natural probable consequence and occurs when the standard actions of an automated system are used inappropriately, e.g., by an under-trained person. The third scenario is direct liability, where both action and intent are present, not likely to happen in medicine as it would require a bad intent from a physician.

Even without employing AI/data mining technologies, the lawsuits in the US medicine still consume   significant amounts of money paid by healthcare providers, physicians, and ultimately the patients through ever increasing premiums.  This should not be the case.  New laws should be enacted regarding usage of medical data to make it much easier to integrate them from different hospitals. We need to remember, however, that the appearance of malpractice might be a fault of simple data-omission or data-transcription errors. The bad outcomes in medicine are not necessarily exclusively the result of negligent physicians or faulty algorithms.

\section{Special status of medicine}

Medical sciences have a special status among other disciplines and one needs to be aware of it when designing data mining methods for diagnostic purposes. Below we review some important aspects of medicine from the data mining point of view.

\noindent $\bullet$ \textbf{Working with human health and life.} Data scientists must be aware of the fact that important information is hidden in the data, one that concerns human health and life. A single mistake may result in a range of complications, including causing undue stress to a patient (incorrect diagnosis, especially for a “fatal” disease), referring a patient for costly and time-consuming additional tests, or even subjecting a patient to an incorrect, potentially harmful, treatment. The worst potential outcome is false negative diagnosis, which is diagnosing a sick patient as healthy. Therefore, data scientists need to keep in mind that even one additional correctly predicted case is worth the time spent on optimizing their models. On the other hand, remembering that perfect diagnosis is a goal that may never be realized: recall that NIH-approved software for automated diagnosis of diabetic retinopathy is only correct 80\% of the time. Physicians are not always correct either.

\noindent $\bullet$ \textbf{Importance of physician interpretation.} Computers cannot and will not replace physicians. They can only simplify their tasks, reduce fatigue and eliminate human-related errors. For young physicians, they are also a great tool for skill improvement. Every diagnostic suggestion made by a machine learning algorithm must be verified by a human expert. This is related to the problem of responsibility as well as the quest for providing the best medical care possible. That is why medical data mining develops clinical decision support systems - not clinical decision-making systems. 

\noindent $\bullet$ \textbf{Need for constant maintenance.} Due to vast amounts of data being generated, as well as high variation among patients, medical data mining algorithms need to be constantly updated with new data collected using better technologies (recall the example about imaging techniques). One cannot assume that a model trained using last year’s data will accurately “diagnose” diseases due to rapid mutations (like in cancer), or antigenic drift (like in influenza), and environmental and social factors changing and impacting human health. 

\noindent $\bullet$ \textbf{Security issues.} Privacy preservation is a crucial aspect of medical data mining. Not only we should avoid leaking personal information and store data in a safe manner, medical data mining should be aware of the potential presence of malicious agents that may extract patient information to advance their agenda. This problem is known as adversarial data mining. It assumes that by providing carefully crafted queries to a data mining system, a malicious user may learn about the data used to train such an algorithm or force it to adapt to new fake information and thus impair its capabilities. 

\section{Conclusions}

This article can be seen as a follow-up on our original paper on uniqueness of medical data mining. It was written because since the time of its publishing rapid advances in technology caused many changes in how medicine is practiced and even taught. We stressed the increasing role of data scientists who analyze big, heterogeneous and dynamic medical data. Such analyses often need to be done in near-real time to provide results when they are needed to be relevant. We also noted the importance of ethical and legal aspects of medical data mining, which are fluid in spite of national and international bodies addressing the issues. Constraints imposed by them determine what types of data may or not be shared, the manner in which they can be analyzed, and sometimes even the conclusions that may be drawn and used in practice. 

\section*{Acknowledgements} 
This paper is dedicated to the late G. William Moore, PhD, MD, recipient of 2007 Lifetime Achievement Award of the Association for Pathology Informatics, a great statistician, pathologist, and colleague (KJC).

\textbf{References}

\bibliographystyle{plain}
\bibliography{refs}



\end{document}